\documentclass[conference, compsocconf]{IEEEtran}

\usepackage{ifpdf}
\usepackage{amssymb,graphicx}
\usepackage{epsfig}
\usepackage{booktabs}
\usepackage[usenames]{color}
\usepackage{multirow}
\usepackage{rotating}

\usepackage{cite}
\usepackage{url}

\newif\ifdraft
\drafttrue
\ifdraft
\newcommand{\hknote}[1]{  {\textcolor{red}  { ***Hartmut: #1 }}}
\newcommand{\mbnote}[1]{  {\textcolor{red} { ***Maciek: #1 }}}
\newcommand{\manote}[1]{   {\textcolor{red}{ ***Matt: #1 }}}
\newcommand{\washnote}[1]{   {\textcolor{red}{ ***Bryce: #1 }}}
\newcommand{\notes}[1]{   {\textcolor{red}   {{\bf NOTE:} #1 }}}
\else
\newcommand{\hknote}[1]{}
\newcommand{\mbnote}[1]{}
\newcommand{\manote}[1]{}
\newcommand{\washnote}[1]{}
\newcommand{\notes}[1]{}
\fi

\begin{document}
\title{Neutron Star Evolutions using Tabulated Equations of State\\ with a New Execution Model}


\author{
\IEEEauthorblockN{Matthew Anderson\IEEEauthorrefmark{1}, Maciej Brodowicz\IEEEauthorrefmark{2}, Hartmut Kaiser\IEEEauthorrefmark{2}\IEEEauthorrefmark{3}, Bryce Adelstein-Lelbach\IEEEauthorrefmark{2}, Thomas Sterling\IEEEauthorrefmark{1}}
\IEEEauthorblockA{\IEEEauthorrefmark{1}Center for Research in Extreme Scale Technology, Indiana University, Bloomington, IN}
\IEEEauthorblockA{\IEEEauthorrefmark{2}Center for Computation and Technology, Louisiana State University, Baton Rouge, LA}
\IEEEauthorblockA{\IEEEauthorrefmark{3}Department of Computer Science, Louisiana State University, Baton Rouge, LA}
\IEEEauthorblockA{andersmw@indiana.edu, maciek@cct.lsu.edu, hkaiser@cct.lsu.edu, blelbach@cct.lsu.edu, tron@indiana.edu}
}

\maketitle

\begin{abstract}
The addition of nuclear and neutrino physics to general relativistic fluid
codes allows for a more realistic description of hot nuclear matter in
neutron star and black hole systems.  This additional microphysics requires
that each processor have access to large tables of data, such as equations
of state,  and in large simulations the memory required to store these
tables locally can become excessive unless an alternative execution
model is used.  In this work we present relativistic fluid evolutions of a neutron star
obtained using a message driven multi-threaded execution model known as ParalleX. 
These neutron star simulations would require substantial memory overhead dedicated 
entirely to the equation of state table if using a more traditional execution model.  
We introduce a ParalleX component based on Futures for accessing large tables of data, 
including out-of-core sized tables, which does not require substantial memory overhead and 
effectively hides any increased network latency.
\end{abstract}


\IEEEpeerreviewmaketitle

%%%%%%%%%%%%%%%%%%%%%%%%%%%%%%%%%%%%%%%%%%%%%%%%%%%%%%%%%%%%%%%%%%%%
%
%   S E C T I O N
%
%%%%%%%%%%%%%%%%%%%%%%%%%%%%%%%%%%%%%%%%%%%%%%%%%%%%%%%%%%%%%%%%%%%%
\section{Introduction}

Future achievements in leading-edge science demand innovations 
in parallel computing models and methods to improve efficiency
 and dramatically increase scalability. One controversial issue 
is the relative value of global address space models and management 
versus more conventional distributed memory structure. This paper 
demonstrates one important use of global address space in the 
context of the advanced ParalleX execution model that has enabled 
simulation improvements to be achieved in the domain of astrophysics 
that is not feasible using conventional practices.

Accurate treatment of hot nuclear matter in astrophysical compact object simulation
is becoming increasingly important in the search for coincident detection of gravitational
radiation and electromagnetic or neutrino emissions originating from the same 
source.  Recent simulations
have shown that the gravitational wave signature from a binary neutron star merger 
or a neutron star--black hole merger
may even reveal important empirical details about the neutron star equation of 
state itself ~\cite{Read2009,Lackey2011}.

The addition of nuclear and neutrino physics to general relativistic fluid
codes allows for a more realistic description of hot nuclear matter in
neutron star and black hole systems.  Unfortunately, most of these and other 
microphysics routines cannot be computed in place; they must be precomputed in a
large table which is then read and interpolated as the relativistic hydrodynamics simulation
proceeds.  Accurate neutron star simulations will increasingly rely upon ever larger
tables of microphysics data which must be read into memory, searched, and interpolated ~\cite{Ott2010}.  
As the memory requirements grow for these tables, approaching and often
exceeding the size of the physical memory of a single node, the need for an alternative
to the traditional MPI-OpenMP hybrid approach is increasingly evident.

In this work we explore the experimental execution model called
ParalleX~\cite{gao, scaling_impaired_apps, tabbal, Dekate2012} 
as a means of addressing the critical computational requirement of dealing with very
large tables containing microphysics.  ParalleX provides many other performance 
benefits to a distributed relativistic hydrodynamics simulation but the
focus of this work will be on matters related to equation of state tables.

ParalleX is a synthesis of complementing semantic constructs delivering a dynamic adaptive 
framework for message-driven multi-threaded computing in a global address space context 
with constraint-based synchronization to exploit locality and manage asynchrony. The result 
is introspective runtime alignment of computing requirements and computing resources while 
permiting asynchronous operation across physically distributed resources. ParalleX has 
been first implemented in the form of the HPX runtime system~\cite{hpx_svn, amr1d}, developed to support the 
semantics and mechanisms comprising ParalleX targeting conventional SMP and commodity 
cluster computing platforms. This experimental software package is developed to test the 
semantics of ParalleX, to measure the overhead costs of software implementation, and 
to provide a prototype of the next generation runtime system for extreme scale applications.

The outline of this paper as follows:  section~\ref{sec:fluid} describes the 
relativistic fluid evolution, initial data, numerical methods, and equation of state
details; section~\ref{sec:hpx} gives a brief overview of the HPX runtime system implementation of ParalleX;
section~\ref{sec:eos} describes how the equation of state is distributed across
nodes and how Futures are used to hide network latency in table access;
section~\ref{sec:results} gives neutron star evolution performance results using 
the Shen equation of state comparing the Futures
approach of accessing the equation of state table with that of reading in the table
for every core (referred to hereafter as the conventional approach); section~\ref{sec:conclusion} gives 
our conclusions and implications for future work.

%%%%%%%%%%%%%%%%%%%%%%%%%%%%%%%%%%%%%%%%%%%%%%%%%%%%%%%%%%%%%%%%%%%%
%
%   S E C T I O N
%
%%%%%%%%%%%%%%%%%%%%%%%%%%%%%%%%%%%%%%%%%%%%%%%%%%%%%%%%%%%%%%%%%%%%
\section{Relativistic Hydrodynamics}
\label{sec:fluid}
The work presented here adopts the flux-conservative formulation of
the relativistic magnetohydrodynamics equations presented in~\cite{Anderson2006}
and includes high-resolution shock capturing (HRSC) methods.  To calculate the
numerical fluxes, we use the Piecewise Parabolic Method (PPM)~\cite{ppm} for 
reconstructing fluid variables. 
The approximate Riemann solver employed is Harten-Lax-van\,Leer-Einfeldt (HLLE)~\cite{hlle}.
While the code is capable of also evolving magnetic fields, no magnetic
fields were added to the initial data at this stage.  Neutron star
simulations presented in section~\ref{sec:results} 
were conducted using the Cowling approximation.

The tabulated equation of state used for generating the initial neutron star and 
all subsequent fluid evolution is the Shen equation of state~\cite{1998NuPhA.637..435S}.
A tabulated Shen equation of state was provided by C. Ott and is available for 
download at ~\cite{stellarcollapse_webpage}.
The publicly available tabulated Shen equation of state used for tests here is
288 MB;  however, an updated Shen equation of state has since been released~\cite{Shen2011}
and we have created a table based on this update which is 5.9 GB in size.  Consequently, 
results from both tables will be presented in the results section.
The neutron star initial data was generated using the Lorene libraries~\cite{lorene_webpage};
the neutron star has a mass of 1.4 solar masses and radius of 15.947 km.  Tables related to 
neutrino transport, including neutrino opacity, were not included in this work
but are part of future work. 

The entire relativistic magnetohydrodynamics solver has been implemented in C++ using the
HPX runtime system for parallelism.  The application has been modularized for future incorporation
into the HPX Adaptive Mesh Refinement (AMR) toolkit although simulations for this work performed all
computations in unigrid.  The next section will give a brief overview of HPX and introduce the concepts
crucial for asynchrony management in tabular access and parallel computation in general.

%%%%%%%%%%%%%%%%%%%%%%%%%%%%%%%%%%%%%%%%%%%%%%%%%%%%%%%%%%%%%%%%%%%%
%
%   S E C T I O N
%
%%%%%%%%%%%%%%%%%%%%%%%%%%%%%%%%%%%%%%%%%%%%%%%%%%%%%%%%%%%%%%%%%%%%
\section{The HPX Runtime System}
\label{sec:hpx}

\begin{figure*}
  \includegraphics[width=0.99\linewidth,height=0.4\textheight]{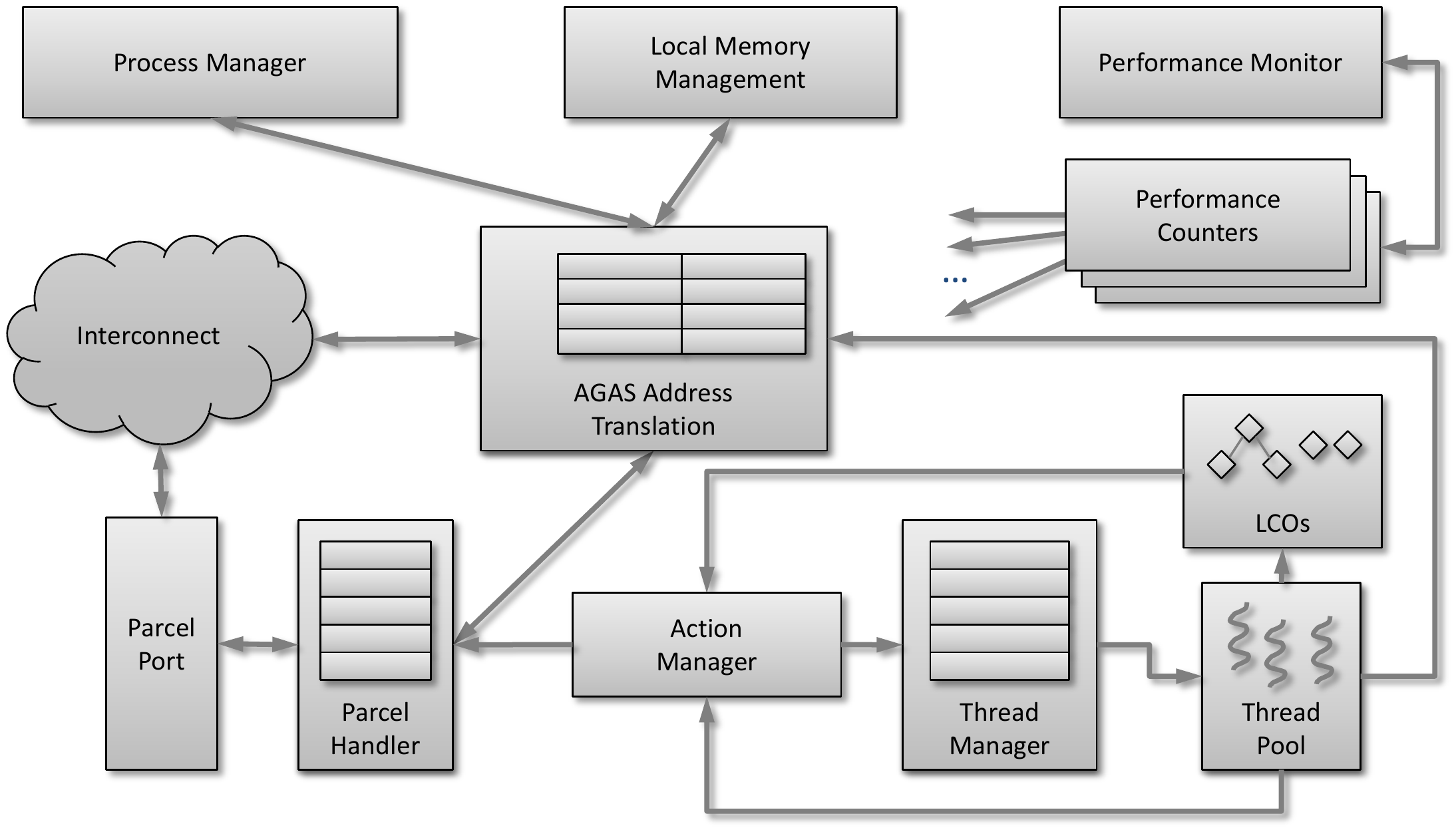}
  \caption{\small{Modular structure of HPX implementation. HPX implements the supporting
     functionality for most of the elements needed for the ParalleX
     model: AGAS (active global address space), parcel port and
     parcel handlers, HPX-threads and thread manager, ParalleX processes, LCOs (local control objects),
     performance counters enabling dynamic and intrinsic system and load
    estimates, and the means of integrating application specific components.}
  }
\label{fig:hpxarch}
\end{figure*}

The C++ prototype runtime implementation of ParalleX is called High Performance ParalleX (HPX).
A walkthrough description of the HPX architecture is found in Figure~\ref{fig:hpxarch}. An incoming parcel 
(delivered over the interconnect) is received by the parcel port. One or more 
parcel handlers are connected to a single parcel port, optionally allowing to distinguish different 
parts of the system as the parcel's final destination. An example for such different destinations 
is to have both normal cores and special hardware (such as a GPGPU) in the same locality (ParalleX
term identifying synchronous domain of computation, such as a single compute node in a cluster). The 
main task of the parcel handler is to buffer incoming parcels for the action manager. The action 
manager decodes the parcel, which contains an action bundled with relevant operands. An action is either a global function
call or a method call on a globally addressable object. The action manager creates a HPX-thread
based on the encoded information.

All HPX-threads are managed by the thread manager, which schedules their execution on one 
of the OS-threads allocated to it. HPX threads are implemented as user level threads, which decreases the 
costs associated with their creation, destruction, and state updates by minimizing the number of
interactions with the OS kernel. HPX creates one worker OS-thread for each available core, 
whose purpose is to carry out the majority of computations in an application. Several scheduling 
policies have been implemented for the thread manager, such as the global queue scheduler, where all
cores pull their work from a single global queue, or the local queue scheduler, where each core
pulls its work from a separate queue. The latter supports work stealing for better
load balancing. In the local scheduler, a queue is created for each of the worker OS-threads.
When a worker thread is searching for work, it first checks its
own queue. If there is no work there, the OS-thread begins to
steal work by searching for work in other queues, first from its own non-uniform memory access (NUMA) domain,
then from cores located on different NUMA domains.

If a possibly remote action has to be executed by a HPX-thread, the action manager 
queries the active global address space (AGAS) to determine whether the target of the action is local or
remote to the locality that the HPX-thread is running on. If the target happens to be local, a new
HPX-thread is created immediately and passed to the thread manager. This thread encapsulates the work (function)
and the corresponding arguments for that action. If the target is remote, the action manager
creates a parcel encoding the action (i.e. the function and its arguments). This parcel is handed
to the parcel handler, which makes sure that it gets sent over the interconnect, causing 
a new HPX-thread to be created at the target locality.

The Active Global Address Space (AGAS) provides global address resolution services
that are used by the parcel port and the action manager. AGAS addresses are 128bit
unique global identifiers (GIDs). AGAS maps these global identifiers to local addresses,
and additionally provides symbolic mappings from strings to GIDs. The local addresses
that GIDs are bound to are typed, providing a degree of protection from type errors.
Any object that has been registered with a GID in AGAS is addressable from all localities
in an instance of the HPX runtime. AGAS also provides a powerful reference counting
system which implements transparent and automatic global garbage collection.

Lightweight Control Objects (LCOs) are the synchronization primitives upon which
HPX applications are built. LCOs provide a means of controlling parallelization
and synchronization of HPX-threads. Semaphores, mutexes and condition
variables~\cite{mutex} are all available in HPX as LCOs. Futures~\cite{future1}
are another type of LCO provided by HPX, and are discussed in greater detail
later in this paper. 

Local memory management, performance counters (a generic monitoring framework),
LCOs and AGAS are all implemented on top of an underlying component framework.
Components are the main building blocks of remotely executable actions and can encapsulate
arbitrary, possibly application specific functionality. Actions are special types which expose the
functionality of a (possibly remote) function. An action can be invoked on a component instance
using only its GID, which allows any locality to
invoke the exposed methods of a component. In the case of the aforementioned components,
the HPX runtime system implements its own functionality in terms of this component framework.
Typically, any application written using HPX extends the set of existing components based on
its requirements.

The relativistic hydrodynamics simulations make use of all the key features of HPX.  The
most crucial feature for contention and network latency hiding in 
tabulated equation of state access is the Future~\cite{future1,future2}.  
The next section will describe the strategy adopted in HPX 
for distributing a large table and hiding network latency.

%%%%%%%%%%%%%%%%%%%%%%%%%%%%%%%%%%%%%%%%%%%%%%%%%%%%%%%%%%%%%%%%%%%%
%
%   S E C T I O N
%
%%%%%%%%%%%%%%%%%%%%%%%%%%%%%%%%%%%%%%%%%%%%%%%%%%%%%%%%%%%%%%%%%%%%
\section{Using the Shen Equation of State Tables}
\label{sec:eos}
% description of Shen EOS and proto-process implementation goes here

The Shen equation of state (EOS) tables of nuclear matter at finite temperature and density 
with various electron fractions within the relativistic mean field (RMF) 
theory are a set of three dimensional data
arrays enabling high precision interpolation of 19 relevant parameters required for 
neutron star simulations.  As noted in section~\ref{sec:fluid}, 
the publicly available Shen equation of
state table is relatively small in size (288\,MB); however, the most recent Shen 
table created for the neutron star evolutions presented here is 5.9\,GB in size.  
Results using both tables will be presented in section~\ref{sec:results}.
In the case of the larger table, loading the whole data set into 
main memory on each locality is not feasible. In conventional MPI based applications the full tables 
would have to be either loaded into each MPI process or a distributed partitioning
scheme would have to be implemented. These options are either not viable or difficult 
to implement using MPI. 

\subsection{Interpolation Technique and Characteristics}
\label{subsec:interpolation}

The values of each of 19 variables describing the Shen EOS are contained in individual
3-D tables stored in memory in a row-major fashion. Single table data are arranged as
samples of a single physical quantity computed at coordinates laying on a regularly spaced grid.
The sizes of each grid vary from $220\times180\times50$ for the smaller 288\,MB set to 
$440\times360\times130$ for the large, 5.9\,GB set. The dimensions correspond to
baryon mass density, temperature, and electron fraction, respectively.

To obtain the value of a variable at an arbitrary point within the 3-D
domain, a $2\times2\times2$ cube of double-precision floating numbers must be accessed.
The result is computed using a simple tri-linear interpolation from these values. 
Our neutron star simulations require only 8 out 19 tabulated quantities, which helps reduce 
the memory pressure. However, even when using a single instance of the smaller set of tables 
on each node to permit the efficent sharing of table data among all cores, the aggregate size of 
accessed data volume still significantly exceeds the combined size of L3 processor caches. 
Furthermore, given that in each interpolation request effectively at most $2\times8=16$ 
out of 64 bytes per cache line (x86 architecture) are used and the coordinate stream generated
by the application is random, the performance of interpolation function is memory bound.

\subsection{The Overhead of Futures}
\label{subsec:futures}

Many HPX applications, including the relativistic hydrodynamics simulation detailed here,
utilize Futures for ease of parallelization and synchronization. In HPX, Futures
are implemented as two types of LCOs, Eager Futures and Lazy Futures, differing in
the evaluation mode of produced value. In Eager Futures the computation providing the result value
is triggered as soon as the Future object is instantiated, whereas for Lazy Futures 
it remains dormant until at least one consumer of the value references it.
The neutron star simulation detailed in this paper makes extensive use of Eager
Futures (we will refer to them simply as Futures for brevity). For this reason, the overheads 
of these constructs are a large factor in
the total overhead of the HPX runtime in our code. In this subsection, we give a
description of Futures, outline a performance test for measuring the overhead of
Futures, and present the results of the test. 

\begin{figure}
  \includegraphics[width=0.99\linewidth]{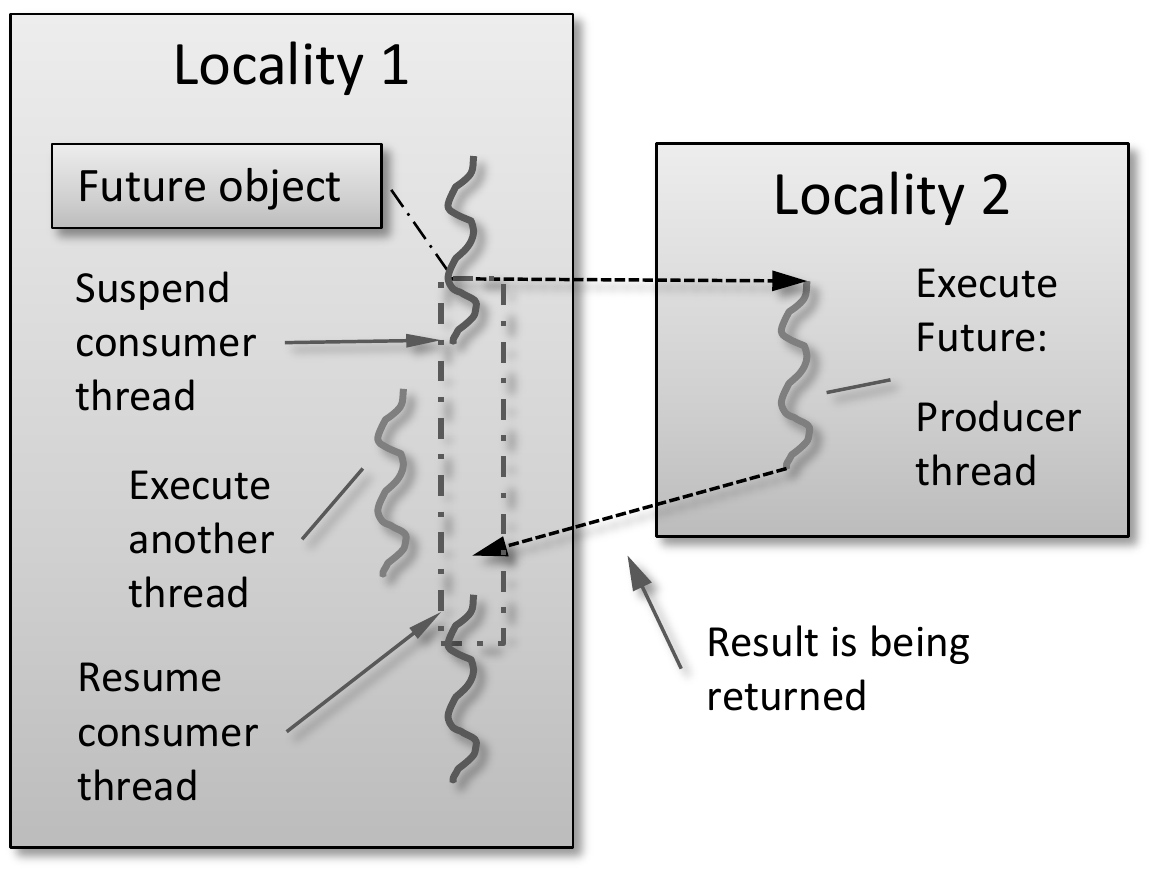}
  \caption{\small{Schematic of a Future execution. At the point of creation of the
     Future, its encapsulated execution is started. The consumer thread is suspended only
     if the result of executing the Future has not returned yet. In this case the core is free to
     execute some other work (here 'another thread') in the meantime. If the result is available,
     the consumer thread continues without interruption to complete execution. The producer
     thread may be executed on the same locality as the consumer thread or on a different
     locality, depending on whether the target data is local or not.}
  }
\label{fig:future_schematics}
\end{figure}

As shown in Figure~\ref{fig:future_schematics}, a Future encapsulates a delayed 
computation. It acts as a proxy for a result initially not known, most of the
time because the computation of the result has not completed yet. The
Future synchronizes the access of this value by optionally suspending
HPX-threads requesting the result until the value is available. When a Future is
created, it spawns a new HPX-thread (either remotely with a parcel or locally
by placing it into the thread queue) which, when run, will execute the
action associated with the Future. The arguments of the action are bound when
the Future is created. 
Once the action has finished executing, a write operation is performed on the
Future. The write operation marks the Future as completed, and
optionally stores data returned by the action. 

When the result of the delayed
computation is needed, a read operation is performed on the Future. If the
Future's action hasn't completed when a read operation is performed on it, the
reader HPX-thread is suspended until the Future is ready. 

\begin{figure}
  \includegraphics[width=0.99\linewidth]{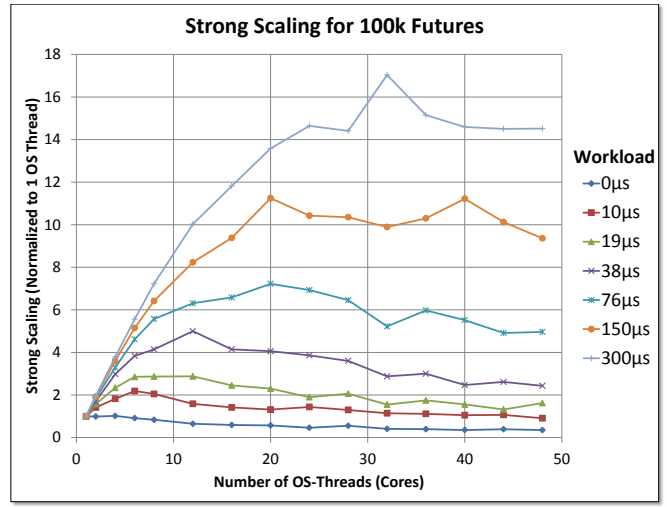}
  \caption{\small{Results of the Future overhead benchmark. In each
    test, 100,000 Futures were invoked, with varying workloads. Each
    data set shows the strong scaling results for a particular workload.}
  }
\label{fig:figure_3}
\end{figure}

Our benchmark for Future overhead created a fixed number of Futures,
each of which had a fixed workload. Then asynchronous read operations were
performed on the Futures until all of the Futures had completed. A high
resolution timer measured the wall-time of the aforementioned operations. The
test was run on an 8-socket HP ProLiant DL785 (each socket sports a 6-core AMD
Opteron 8431) with 96 GB of RAM (533 MHz DDR2). Trials were done with
varying workloads and OS-threads. Five runs were performed for each combination of
the parameters and the results were averaged to produce a final dataset. The
numbers are presented in Figure~\ref{fig:figure_3}.

On the locality we used for this benchmark, the amortized overhead of a 
 Future is approximately \emph{17 microseconds}. This overhead includes the 
time required to create the Future instance, start the evaluator thread,
perform synchronization with the accessors, and destruct the Future object.
This number was extrapolated
from the data presented in Figure~\ref{fig:figure_3}. We multiplied
the workload by the number of Futures used in each run, and then subtracted
that from the average wall-time of the trial. We divided that number by the 
number of  Futures invoked in the trial to get the overhead per Future for
each set of parameters. 

\begin{eqnarray*}
& \mbox{\textit{overhead}} = \frac{\mbox{\textit{avg. wall-time}}\,-\,(\mbox{\textit{workload}}\,*\,\mbox{\textit{futures invoked}})}{\mbox{\textit{futures invoked}}}
\end{eqnarray*}
 
The scaling results in Figure~\ref{fig:figure_3} call for some
explanation. The parabolic curves are formed primarily by contention in the
thread queue scheduler. As the number of OS-threads is increased, the contention
on the thread queue scheduler also increases, due to a higher number of
concurrent searches for available work. This increased contention occurs in both
global queue schedulers (where all OS-threads poll the same work queue, and must
obtain exclusive access to said queue for some period of time) and to some
extent in local queue schedulers (where work stealing occurs, which causes queue
contention, albeit to a lesser degree than in the global queue scheduler). As
we increase the workload in each Future, OS-threads spend more time
executing the workloads and less time searching for more work. This decreases 
contention on the queues. Adding a new OS-thread is beneficial as long as the
contention overhead that it causes is not greater than the parallel speedup that
it provides. 

\subsection{The Overhead of the Shen EOS Table Partitioning}
\label{subsec:sheneos}

\begin{figure}
  \includegraphics[width=0.99\linewidth]{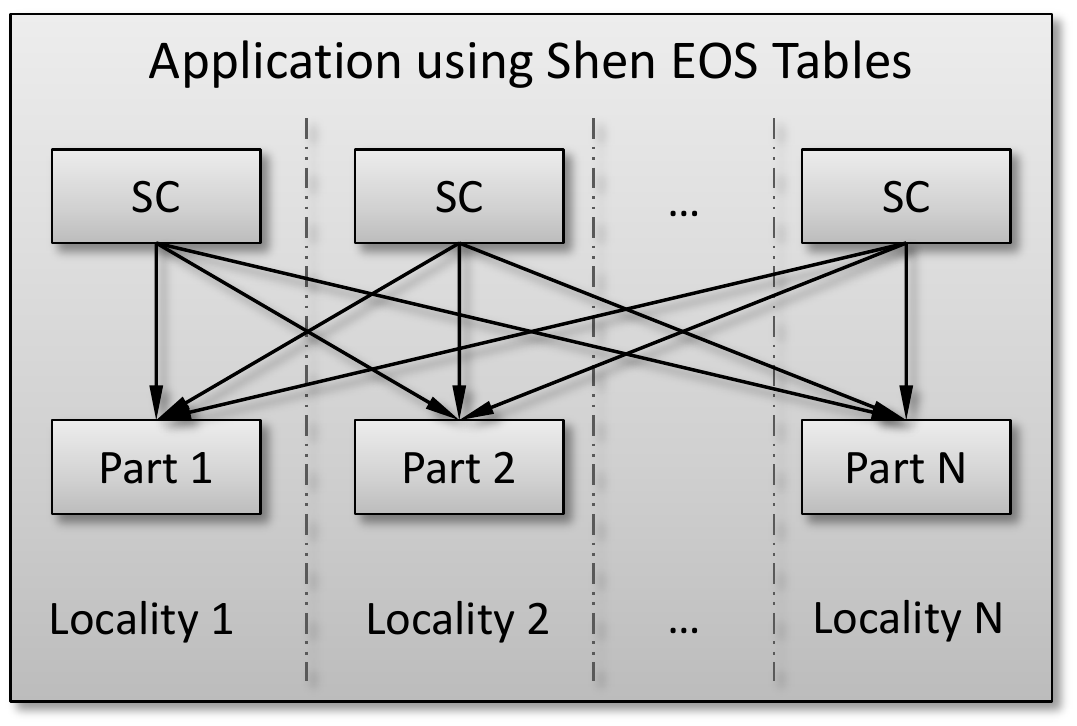}
  \caption{\small{Schematic of an application using the distributed partitioned Shen equation of state (EOS)
    tables. Each locality has a Shen EOS client side object (SC) allowing to transparently access all of the table data.
    At the same time the Shen EOS table data is partitioned into chunks of approximately equal size, each of which is loaded into
    the main memory of one of the localities (Part 1~...~Part N) thus lessening the required memory footprint
    for each of the localities.}
  }
\label{fig:sheneos}
\end{figure}

We created an HPX component encapsulating the minimally overlapping partitioning (ghost zone of single element 
width) and distribution of the Shen EOS tables to all available localities, thus reducing the required 
memory footprint per locality~\cite{sheneos_svn}. A special client side object ensures the transparent 
dispatching of interpolation requests to the appropriate partition corresponding to 
the locality holding the required part of the tables (see Figure~\ref{fig:sheneos}).
The client side object exposes a simple API for easy programmability.

% Shen EOS test results
The second part of this section describes the setup and results of the
measurements we performed in order to estimate the overheads introduced
by distributing the Shen EOS tables across all localities. To evaluate the
scalability and associated overheads of the distributed implementation
of the Shen EOS tables, a number of tests have been
performed, all of them with a fixed number of total data accesses
(measuring strong scaling). The tests have been run on a different
number of localities and with varying numbers of OS-threads per locality.
The current HPX implementation supports only a centralized AGAS server
that may be invoked in two configurations: either as a standalone task
on a dedicated locality or as a part of one of the user application tasks.
Our tests used a standalone AGAS server, firstly to avoid interfering with
the user workload and secondly to eliminate the
generation of asymmetric AGAS traffic on localities hosting data
tables. Unlike the client applications, the AGAS server used a
fixed number of OS-threads throughout the testing to ensure that
sufficient processing resources are available to the incoming
resolution requests.

The tests were performed on a small heterogeneous cluster. The cluster consists
of 18 localities (excluding the head node) connected by Gigabit Ethernet network. Two of
the machines are 8-socket HP ProLiant DL785s, with 6-core AMD Opteron 8431s and 96
GB of RAM (533 MHz DDR2). The other 16 localities are single-socket HP ProLiant DL120s,
with Intel Xeon X3430s and 4 GB of RAM (1332 MHz DDR3). All machines run
x86-64 Debian Linux. Torque PBS was used to run multi-locality tests. 

\begin{figure}
  \includegraphics[width=0.99\linewidth]{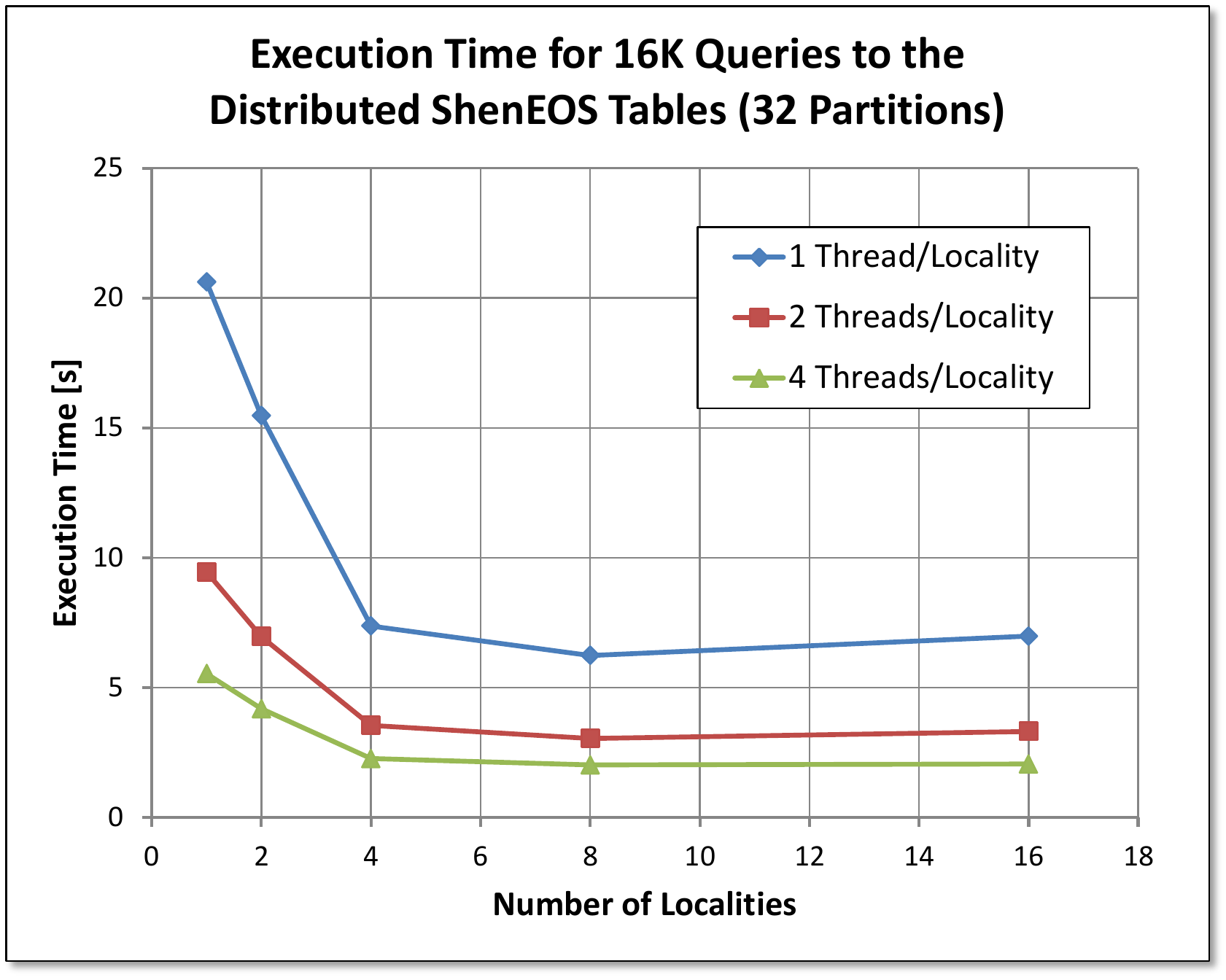}
  \caption{\small{Scaling of the execution time for 16K separate non-bulk data interpolation
    queries to the Shen EOS tables distributed across 32 partitions measured
   for different number of localities and varying number of OS-threads per locality.}
  }
\label{fig:sheneos_execution_time}
\end{figure}

Figure~\ref{fig:sheneos_execution_time} shows the execution times
collected for the data access phase with a special test application executed 
on up to 16 localities and 1, 2, and 4 OS-threads per locality. The total number of
distributed partitions was fixed at 32 to preserve the AGAS traffic
pattern when run on a different number of localities; all partitions
were uniformly distributed across the test localities. The number of 
separate, non-bulk queries to the distributed Shen EOS partitions was set to a fixed 
number of 16K. Each of these queries created a Future encapsulating
the whole operation of sending the request to the remote partition,
schedule and execute a HPX-thread, perform the interpolation based on
the supplied arguments for the Shen EOS data, sending back the resulting
values to the requesting HPX-thread, and resuming the HPX-thread which was
suspended by the Future in order to wait for the results to come back.

% Notes:
% - look like 1 AGAS server is sufficient to handle at least 16-locality
%   heavy AGAS traffic (4 cores each) on GigE; interpolation overhead
%   is not significant

The graph demonstrates that the overhead of distributed table
implementation does not increase significantly over the entire range
of available localities.  While the scaling is much better when the
number of localities remains small (up to 4), the overall time
required to service the full 16K data lookup requests remains roughly
constant.
The test application itself does not execute any work besides 
querying and interpolating the distributed tables, which does not leave much room
to overlap the significant network traffic generated with useful
computation. This causes the scaling to flatten out beyond
8 localities. Using the distributed tables in real applications doing much more
work will allow to further amortize the introduced network overheads. 
The results also imply that a single AGAS server is quite capable of servicing 
at least 16 client localities, especially considering the intensity of request
traffic over Ethernet interconnect deployed in out testbed.  We plan
to further evaluate this aspect of distributed table implementation
using faster interconnect networks, such as Infiniband.

%%%%%%%%%%%%%%%%%%%%%%%%%%%%%%%%%%%%%%%%%%%%%%%%%%%%%%%%%%%%%%%%%%%%
%
%   S E C T I O N
%
%%%%%%%%%%%%%%%%%%%%%%%%%%%%%%%%%%%%%%%%%%%%%%%%%%%%%%%%%%%%%%%%%%%%
\section{Results}
\label{sec:results}

\begin{figure}
  \includegraphics[width=0.99\linewidth]{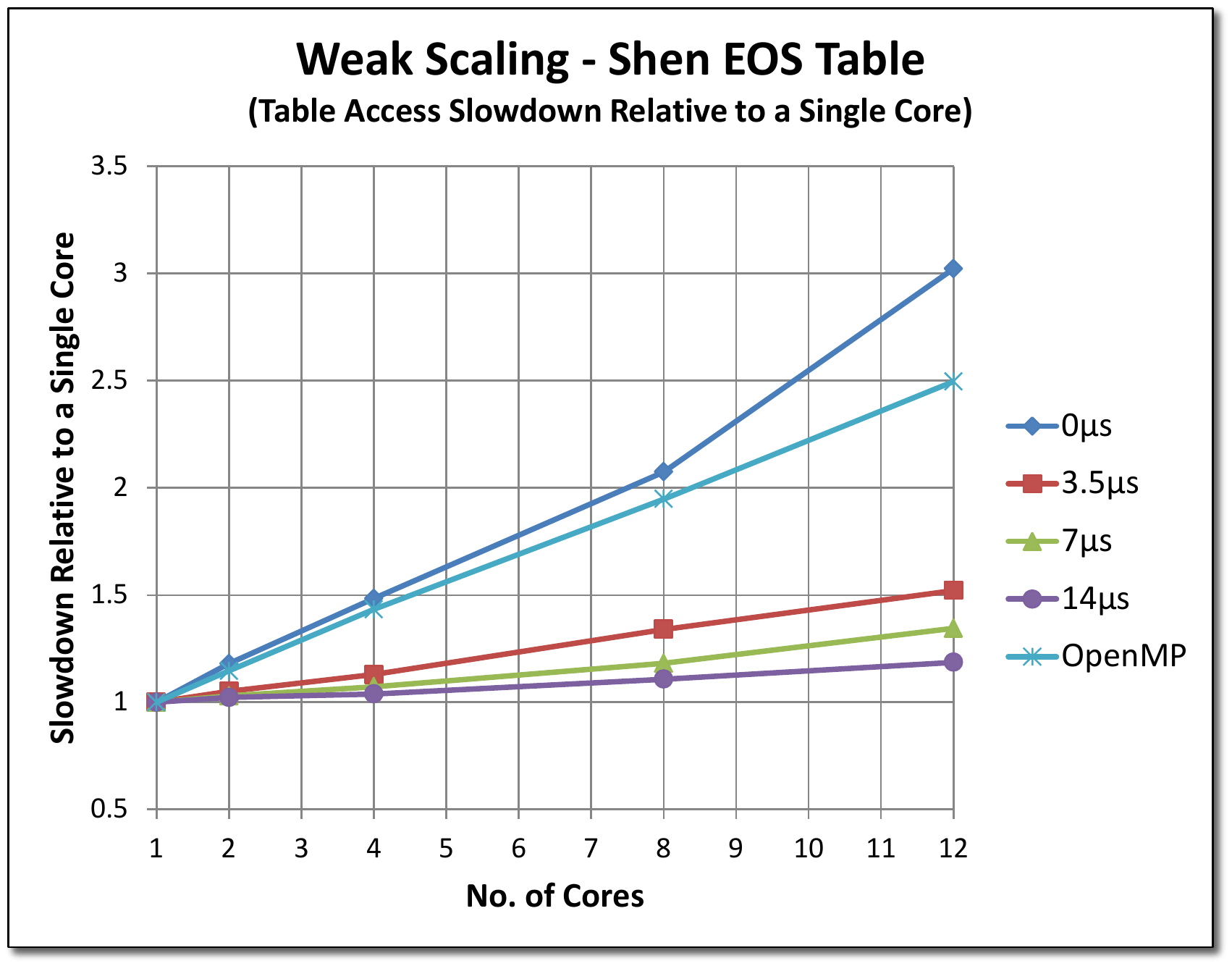}
  \caption{\small{The relative slowdown in table access when run across
various numbers of cores on a shared memory machine.  For comparison, results 
using OpenMP are also provided;  all other results use HPX for table access.  
When no additional work is overlapped with the table access,
the slowdown relative to that seen on a single core can reach as high as a 
factor of 3;
however, when other workload is overlapped with the table access, the contention
in table access is increasingly amortized.  These results
used the smaller (288 MB) table.
  }}
\label{fig:ws_shen}
\end{figure}

Accessing a single, potentially distributed Shen equation of state table 
using multiple threads
for converting conservative variables to primitives and vice versa as
required for the flux-conservative HRSC method results in a slowdown
when compared with using multiple independent copies of the table.  There
is also some additional overhead in using futures in the tabular access.
In Figure~\ref{fig:ws_shen} the table access slowdown relative to single
core on a shared memory machine is presented.  The results are a weak scaling test where the results
for each workload have been normalized to the corresponding 
single core performance.  In this test, each core accesses and interpolates 
64k unique values in the table as a single bulk operation.  Using HPX on a single core of an 
Intel Xeon X5660 processor, this test takes 0.0728 seconds;
for comparison, using the Fortran codes 
provided at~\cite{stellarcollapse_webpage}   
access and interpolation of the exact same 64k values takes 0.0549 seconds,
reflecting the increased overhead in using HPX.  As the number of cores 
accessing the same table increases, the access performance degrades.
The primary reason for that is the competition of hardware threads executing
on the same processor for access to memory, since most of the interpolation requests 
cannot be satisfied solely from processor caches (for the machine used in test, the 
ratio of utilized fraction of EOS dataset to the aggregate size of L3 caches was about 5).
This is compounded even further by the fact that accesses are sparse and random in nature, and 
therefore result in decrease of the effective memory bandwidth. When 
no additional work is overlapped 
with the table access, the table access slowdown relative to a single core
can reach as high as a factor of 3, an unacceptably high number for neutron
star simulations.  However, by overlapping work with the table access futures,
the slowdown relative to a single core becomes much more reasonable.

For relatively small tables like the 288 MB table explored here, 
the memory cost of reading in a 
table for each core might be a manageable strategy 
in order to avoid any higher overhead costs associated with sharing the table 
using futures.  
For very large tables, however, there is
no other viable option: the table would have to be distributed across
several nodes and shared using  futures.  Using the recently released
improved Shen equation of state~\cite{Shen2011},
we have created Shen tables 5.9 GB in size for neutron star simulations.
The increased table resolution contributes to improving the
 accuracy and robustness of neutron star evolutions
as well as includes the most recently improved RMF results.

\begin{figure}
  \includegraphics[width=0.99\linewidth]{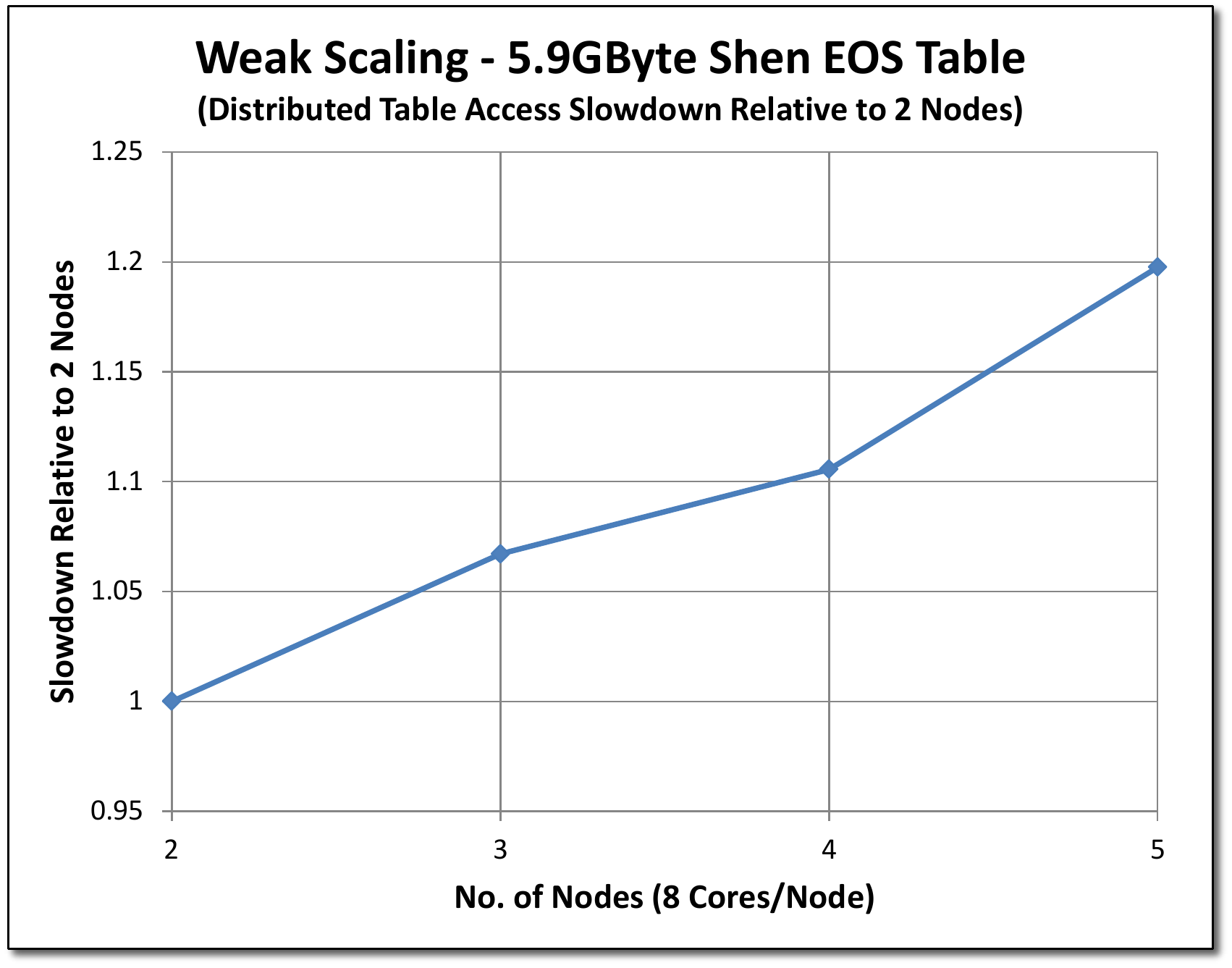}
  \caption{\small{The relative slowdown in table access when run across
various numbers of nodes where on each node 8 cores are used.  This distributed test used 
a 5.9 GB table based on the latest Shen equation of state~\cite{Shen2011}.    
The interconnect was gigabit Ethernet; the workload provided was only that
of table access and interpolation. This shows that the cost of distributing a very large 
table over a number of localities is acceptable.
  }}
\label{fig:ws_distrshen}
\end{figure}

In Figure~\ref{fig:ws_distrshen} the table access slowdown relative for the
larger table is presented.  This figure presents a weak scaling table access
test where each core accesses and interpolates 64K different table queries.  
Because of the size of the table and the need for
memory dedicated to the fluid field meshes and evolution, 
this table has to be distributed.  The distributed results presented 
have been normalized 
to the corresponding two node performance.  We use 8 cores in each node, 
consisting of two quad-core
Intel Nehalem (2.8 GHz) processors with gigabit ethernet interconnect.  The 
only workload provided was that of table access and interpolation; 
no additional workload was added.  For this
larger table, the relative slowdown in distributing the table is, at worst,
a factor of 1.2 if distributed across 5 nodes with a benefit of 
saving enormous amounts of memory.
There is no toolkit or routine at present based on a different execution model 
against which to compare this result.

\begin{figure}
  \includegraphics[width=0.99\linewidth]{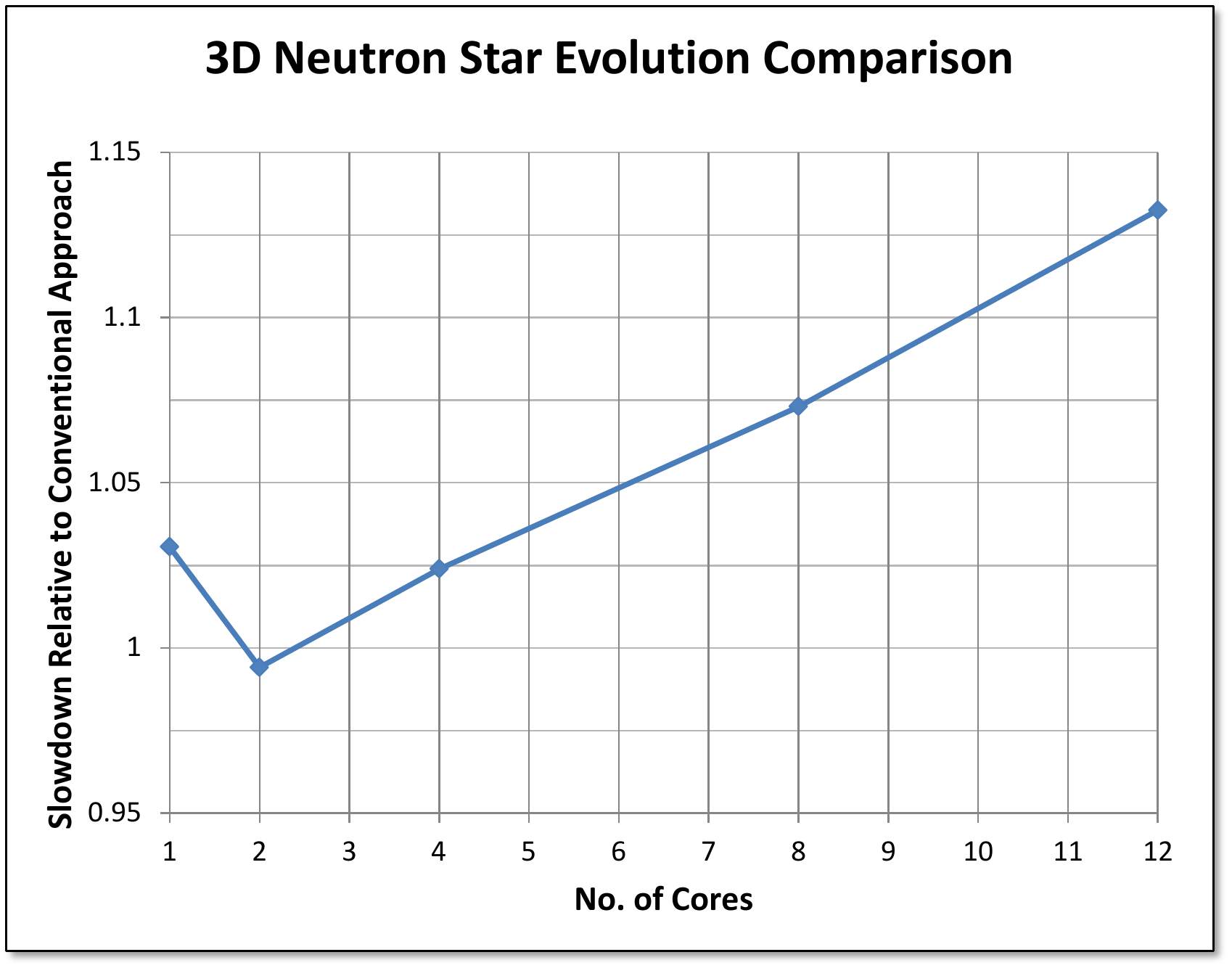}
  \caption{\small{The relative slowdown in running a 3-D Neutron star
evolution with a finite temperature equation of state for 10 iterations 
on a shared memory machine
comparing Futures based table access with the traditional approach - reading in the table
for each core.
The table used in the comparison is the smaller (288 MB) Shen table.  The
simulation was unigrid with $50^3$ points across the computational domain. 
We find the relative slowdown in using Futures based table access 
compared to reading in the table on multiple cores to be 
extremely minimal -- at worse a factor of 1.13 when table access 
is shared across 12 cores -- with a considerable savings in memory.
These results resemble the 14 $\mu$s workload line seen in 
Figure~\ref{fig:ws_shen}.
  }}
\label{fig:ns}
\end{figure}

In Figure~\ref{fig:ns} we evolve a neutron star with the Shen equation of state
on a shared memory machine (Intel Xeon X5660, 12 cores)
for 10 iterations using a grid with $50^3$ points across the computational
domain and compare performance between the Futures based table access approach
and the conventional approach of reading in a separate table for each core.
By necessity this comparison had to use the smaller (288 MB) Shen table 
because using the larger table would have exceeded the memory available when
testing the conventional approach.  The Futures based table access performs
extremely well compared to the conventional approach with negligible
slowdown on up to 4 cores.  At worst, when the table is shared across 12 cores,
the slowdown is a factor of 1.13, consistent with the numbers presented
in Figure~\ref{fig:ws_shen} for the 14$\mu$s workload case. 

%%%%%%%%%%%%%%%%%%%%%%%%%%%%%%%%%%%%%%%%%%%%%%%%%%%%%%%%%%%%%%%%%%%%
%
%   S E C T I O N
%
%%%%%%%%%%%%%%%%%%%%%%%%%%%%%%%%%%%%%%%%%%%%%%%%%%%%%%%%%%%%%%%%%%%%
\section{Conclusion}
\label{sec:conclusion}
We have examined finite temperature tabulated equation of state access 
in the context of neutron star simulations using a relatively new execution
model called ParalleX.  Using Futures to manage asynchrony, amortize contention,
and hide network latency, we have presented a strategy for performing neutron star evolutions
using extremely large tabulated equations of state with minimal performance and
memory cost.  Using the Futures based partitioned Shen EOS table access the slowdown compared to 
the conventional way of accessing these tables is less than $\sim$15\%, and often much less
than that. This added cost is justifiable, since 
as larger tables become available in simulation efforts, 
astrophysics simulations can then achieve a more realistic description of hot nuclear matter
and incorporate more microphysics, including neutrino transport. Managing large tables
in this asynchronous way would be difficult to implement when 
using conventional programming models, such as MPI.

Several key improvements to the results presented here are currently underway.  
As the HPX runtime system becomes NUMA aware, much of the memory contention observed
here in both OpenMP and HPX runs can be eliminated~\cite{heller}.  Ways to reduce the Futures
overhead reported in Fig.~\ref{fig:figure_3} even further 
are currently under investigation.  All distributed runs presented here used gigabit
Ethernet interconnect; however, HPX support for the native Verbs interface for Infiniband
is also underway. Hardware support for AGAS translation, whose first-cut implementation
could utilize FPGA (Field-Programmable Gate Array) technology, 
promises to reduce key overheads, both in execution and storage, related to the software 
implementation.  OpenCL support
via percolation in HPX is also under development and could subtantially impact the
capability to perform neutrino transport in neutron star simulations. 

As a final note, we point out that switching from a message passing to message driven
style computation for neutron star simulations has significant performance impacts beyond 
just those discussed in this paper involving the finite temperature equation of state tables.
While those improvements are outside the scope of this work, the key concepts of 
managing asynchrony, amortizing contention, and hiding network latency make a significant
positive impact in the scalability of neutron star simulations. 

%%%%%%%%%%%%%%%%%%%%%%%%%%%%%%%%%%%%%%%%%%%%%%%%%%%%%%%%%%%%%%%%%%%%
%
%   S E C T I O N
\section*{Acknowledgments}
We thank Steven Brandt, Luis Lehner, and David Neilsen for stimulating discussions. 
We also thank Lloyd Brown for his technical assistance.  Computations were done at
Brigham Young University (Marylou5), Lousiana State University (Hermione), and 
Indiana University.  This work was supported by the NSF under grants CNS-1117470
and CCF-1142905.


\bibliographystyle{IEEEtran}
\bibliography{IEEEabrv,pxBib}

\end{document}